\documentclass[twocolumn,showpacs,nofootinbib,10pt]{revtex4-1}
\usepackage[colorlinks,hyperindex]{hyperref}
\usepackage{epsfig}
\usepackage{float,upgreek}
\usepackage[colorlinks,hyperindex]{hyperref}
\RequirePackage{amssymb,amsmath}
\usepackage[latin1]{inputenc}
\usepackage{graphicx}
\usepackage{indentfirst}
\usepackage{color}
\begin{document}

\title{The nuclear configurational entropy impact parameter dependence in the Color-Glass Condensate}
\author{G. Karapetyan}
\affiliation{Centro de Ciencias Naturais e Humanas, Universidade Federal do ABC - UFABC\\ 09210-580, Santo Andre, Brazil.}
\email{gayane.karapetyan@ufabc.edu.br}

\begin{abstract}
The impact parameter (b) dependence on the saturation scale, in the framework of the Color Glass Condensate (b-CGC)
dipole model, is investigated from the configurational point of view.
During the calculations and analysis of the quantum nuclear states,  the critical points of stability in the configurational entropy setup
are computed, matching the experimental parameters that define the onset of the quantum regime in the b-CGC 
in the literature with very good accuracy. 
This new approach is crucial and important for understanding the stability of quantum systems in study of deep inelastic scattering processes.\end{abstract}
\pacs{89.70.Cf,24.85.+p}

\maketitle

\section{Introduction}

The nuclear configurational entropy \cite{Gaya} takes into the cross-section as a natural localized, square-integrable functions, driving the configurational entropy setup in nuclear physics. This concept has been already applied to gauge dualities involving quantum chromodynamics (QCD) models in Refs. 
 \cite{Bernardini:2016hvx,Bernardini:2016qit} for studying the stability of mesons and scalar glueballs. The {nuclear configurational entropy}  
is based upon the information entropy, implemented by the recently 
introduced configurational entropy \cite{Gleiser:2011di,Gleiser:2015rwa,Gleiser:2012tu,Sowinski:2015cfa}, for spatially-localized physical systems.
As the the cross section of any nuclear reaction is  spatially-localized and employed to characterize the probability that any reaction occurs,
the analysis of the shape complexity of classical field configurations can be implemented in the context of cross sections. 
In nuclear physics one measures the cross section values for the systems with a finite spatial extent. 
In order to describe the system, thus, one needs as much  information, as the higher the informational entropy is.
A brief analysis between the configurational entropy  lattice approach and statistical mechanics has been paved in Ref. \cite{Bernardini:2016hvx}.  In a similar context, the Hawking-Page transition
was analysed \cite{ads} and {Bose-Einstein condensates have been studied} \cite{Casadio:2016aum} with interesting results. 

 The strong interaction at high energy regimes has been studied in both experiment and
theory.  QCD is a theory of strong interactions that is able to explain most of their underlying experimental results to remarkable accuracy.
Another important source of information is {numerical lattice calculation}, which produces correct solution to the QCD field equations.
However, a lot of questions still is open and the lattice data still need
to be interpreted in order to find the mechanisms and principles that lead to these {numerical results}.
In this case, the Shannon based informational entropy can shed some light onto the principal results based on
the lattice QCD setup and introduces a quantitative theoretical apparatus for studying the instability of high-excited
nuclear system. In this work, the cross sections  will be  analysed  with regard to some principal points of the associated informational entropy.
{Our focus in this letter is applying} the Color Glass Condensate
to nuclear collisions.

Colliding heavy nuclei with {each other makes} QCD to predict that a
new state of matter, called the quark-qluon plasma (QGP), is then formed \cite{Song:2008hj}. 
The most precise measurements of the quark
and gluon structure of the proton come from the HERA particle accelerator,
which collided electrons and positrons with protons.
In these experiments, the proton can look very
differently, depending on at which scale it is measured. When the proton
structure is probed with a photon that has a long wavelength compared with
the proton size, a charged particle with electric charge $+e$ is seen. The
inner structure of the proton becomes visible when the photon wavelength is
decreased to the order of the proton radius. First, one observes three
valence quarks having a fractional electric charge and carrying a fraction $\sim$ 1/3
of the proton longitudinal momentum, viewed in the frame where the proton
energy is very large.
When the wavelength of the photon decreases more, a richer structure becomes visible. The photon starts to
see a large number of sea quarks and antiquarks that carry a small fraction
of the proton longitudinal momentum, denoted by Bjorken-$x$ \cite{kkt,Carvalho,cgc,Carvalho:2008zzb}.
 These quarks
originate from gluon splittings to quark-antiquark pairs. 
 When the proton structure is
measured at smaller $x$, more sea quarks and gluons are seen. 
The mechanism of the reaction based on QCD, can be investigated by the dependence of the total hadronic cross section on the energy, as well as the study of the appropriate role of partons during the interaction with high energy hadrons.
Such important questions which have arose since the early days of the strong interactions,
as, for instance, the behavior of cross sections at high energies, or the universality
of hadronic interactions, as well as the nature of multi-particle
production, have acquired a new vision.
In the KKT model \cite{kkt}, it was assumed that that the gluons inside particles are seen to other particles as a gluon wall, that describes the Color Glass Condensate itself.
In other words, gluons from the inside have a high density distribution. Increasing the energy increases the momentum states that are occupied by
the gluons, forcing a weaker coupling among the gluons \cite{cgc,kkt,Carvalho}.
This leads to the gluon saturation effect, which corresponds
to a multiparticle Bose condensate state. 

QCD at high energies can be described as a many-body theory of partons which
are weakly coupled albeit non-perturbative due to the large number of
partons.
Such a system is called a Color Glass Condensate (CGC) and describes
an effective perturbative weak-coupling field theory approach in the small-$x$ regime of QCD.
The initial conditions
for high energy collisions are determined by the free
partons  in the wave functions of the colliding
nuclei.

Partons represent the localized-energy systems, they can be considered as the spatially-coherent field configurations in an
informational entropic context, together with the Shannon entropy of information theory \cite{Gleiser:2012tu}.
The critical points of the configurational entropy correspond to the onset of instability of a spatially-bound configuration. 
In a considerable amount of studies, the configurational entropy has been applied to analyze aspects in a
variety of models, which comprise the higher spin mesons and glueballs stability \cite{Bernardini:2016hvx, Bernardini:2016qit}
as well as  Bose-Einstein condensates of long-wavelength gravitons, which describes black holes. \cite{Casadio:2016aum}.
The physical systems, that were studied in above mentioned works, had configurations of classical fields and have been successfully analyzed from the view of critical points, including the derivation of the Higgs mass \cite{Alves:2017ljt,Alves:2014ksa}. 

It is interesting to  analyze the configurational entropy from the point of view of spatially-localized reaction cross sections as \cite{Gaya} 
\begin{equation}\label{34}
{\boldsymbol\upsigma} (\vec{k})=\frac{1}{2\pi}\! \iint_{-\infty}^\infty  \!\upsigma(\vec{r})\,e^{i\vec{k} \cdot\vec{r}} d \vec{r}\,\,.
\end{equation}  Then the modal fraction can be expressed as \cite{Gleiser:2011di,Gleiser:2015rwa,Gleiser:2012tu,Sowinski:2015cfa}:
\begin{equation}\label{modall}
f_{\boldsymbol\upsigma}(\vec{k})=\frac{\vert\,\boldsymbol\upsigma(\vec{k})\,\vert^2}{\iint_{-\infty}^\infty\vert\,{}{\boldsymbol\upsigma(\vec{k})\vert\,^2}\,d\vec{k}}.
\end{equation}
 Finally, one can define the configurational entropy analogous to what has been defined for the energy density in Ref. \cite{Gleiser:2012tu}, but this time for the cross section, as
\begin{equation}\label{333}
S_c\left[f\right] \,= \, - \iint_{-\infty}^\infty \dot{f}_{\boldsymbol\upsigma}(\vec{k} ) \log  \dot{f}_{\boldsymbol\upsigma} (\vec{k})  d \vec{k}, 
\end{equation}
where $\dot{f}(\vec{k}) = {f}(\vec{k})/f_{\rm max}({\vec{k}})$.  
Critical points of the configurational entropy  imply that the system has informational entropy
that is critical with respect to the maximal entropy $f_{\rm max}({\vec{k}})$, corresponding to
more dominant states \cite{Bernardini:2016hvx,ads,Casadio:2016aum}.

\section{The configurational entropy and KKT model}

Study of deep inelastic scattering (DIS) reactions and exclusive
diffractive processes of leptons on protons and/or nuclei, such as, for
instance, exclusive vector mesons production and virtual Compton
scattering at small-$x$, is turned toward understanding the
mechanism of QCD. Many interesting questions which have arisen since
the beginning of the investigation of strong interactions, such as the
behavior of cross sections at high energies, the universality of
hadronic interactions at high energies, have acquired fresh vigor.
 In the CGC approximation, one of the
important ingredients for particle production is the universal dipole
amplitude, represented by the imaginary part of the quark-antiquark
scattering amplitude on a target. 

The impact parameter (b) dependence of
the dipole amplitude is essential for understanding exclusive
diffractive processes in the CGC or the color dipole approach.
Therefore, in this work we will briefly represent a simple dipole 
model, b-CGC, that incorporates all known properties of the gluon
saturation, and the impact parameter dependence of dipole amplitude
\cite{watt-bcgc}. The b-CGC model has been applied to various
reactions, as  deep inelastic scattering and diffractive processes \cite{watt-bcgc} and
proton-nucleus \cite{pa-R} collisions at RHIC and the LHC.

At small $x$, the  deep inelastic scattering is characterized by the fluctuation of the virtual
photon $\gamma^{\star}$ into a quark-antiquark pair $q\bar{q}$ with
size $r$, which then scatters off the hadronic or nuclear target via
gluon exchanges. Then, the total deep inelastic cross section,
$\gamma^{\star}p$, for a given Bjorken  $x$ and virtuality $Q^2$
will be expressed as \cite{ni}: 
\begin{eqnarray}
 \label{CA4}
  \upsigma_{L,T}^{\gamma^*p}(Q^2,x) \!\!\!&\!\!=\!\! &\!\!\!2\sum_f \int\!\,d^2\vec{r}\int\!\,d^2\vec{b}\int_0^1 dz\, |\Psi_{L,T}^{(f)}(r,z,m_f;Q^2)|^2
\nonumber\\&&  \,\times\,\,\mathcal{N}\left(x,r,b\right),
\end{eqnarray}
where $z$ is the fraction of the light-front momentum of the virtual
photon carried by the quark, $m_f$ is the quark mass, and 
$\mathcal{N}\left(x,r,b\right)$ is the imaginary part of the forward
$q\bar{q}$ dipole-proton scattering amplitude with dipole size $r$
and impact parameter $b$. The form of light front wave function
$\Psi_{L,T}^{(f)}$, for $\gamma^{\star}$ fluctuations into
$q\bar{q}$, was taken from Ref.\,\cite{gbw}. The subscript $L,T$  in
Eq. (\ref{CA5}) denotes the longitudinal and transverse polarizations of the
virtual photon. In Ref. \cite{IIM} it was suggested a simple dipole
model, which links two limiting behavior  of Eq.
(\ref{CA5}), namely in the vicinity of the saturation line for small dipole
sizes, $r \ll1/Q_s$, and for the case of deep inside the saturation
region for larger dipoles, $r \gg1/Q_s$. Such a model is historically
called CGC dipole model, in which the color dipole-proton amplitude
is given by:
\begin{eqnarray}
 \label{CA5}
N\, (x, r, b)=\left\{\begin{array}{l}N_0(\frac{r Q_s}{2})\,^{2\gamma_{eff}},\,\,\,r Q_s\,\leq\,2\,,\\ \\
1\!-\!\exp(-\mathcal{A} \log^2(\mathcal{B} r Q_s)),\,\,\,\,\ rQ_s\,>\,2\,,\end{array}
\right.
\end{eqnarray}
where the effective anomalous dimension is expressed  by formula
\begin{eqnarray}
 \label{g-eff}
\gamma_{eff}=\gamma_s\,\,+\,\,\frac{1}{\kappa \lambda Y}\log\left(\,\frac{2}{r Q_s}\right)\,,
\end{eqnarray}
where $Y=\log(1/x)$ and $\kappa= \chi''(\gamma_s)/\chi'(\gamma_s)$, with $\chi$ being the
characteristic function. The scale $Q_s$ in Eqs.~(\ref{CA5},\ref{g-eff}), is generally called the saturation scale. In the CGC dipole model, the scale $Q_s$ is given by following expression:
\begin{eqnarray}
 \label{qs}
Q_s\mapsto Q_s(x)=\left(\, \frac{x_0}{x}\right)\,^{\frac{\lambda}{2}} \text{GeV}.
\end{eqnarray}
The parameters $\mathcal{A}$ and
$\mathcal{B}$ in Eq.(\ref{CA5}), are determined from the matching of the dipole amplitude and
its logarithmic derivatives at $rQ_s=2$:
\begin{eqnarray}
\mathcal{A} =-\frac{N_0^2\gamma_s^2}{(1-N_0)^2\log(1-N_0)}, \hspace{0.3cm} \mathcal{B} =\frac{1}{2}(1-N_0)^{-\frac{1-N_0}{N_0\gamma_s}}.
 \label{AB}
\end{eqnarray}

The amplitude is considered to be independent from the impact parameter
in the Color Glass Condensate model. Thus, the integral over
impact parameter in Eq. \eqref{CA4} can be considered as a normalization
factor  $\upsigma_0=2\int d^2b$, and is determined by a fit to data.
Therefore, the total dipole cross section will be estimated as
$\upsigma_{q\bar{q}}=\upsigma_0 \mathcal{N} (x,r)$. The parameters
$\kappa=9.9$  and $N_0=0.7$ are fixed \cite{watt-bcgc, IIM}, and the
other four parameters, namely $\gamma_s, x_0, \lambda, \upsigma_0$,
are obtained by a fit to the HERA data via a $\chi^2$ minimization
procedure.

In the  so-called b-CGC model \cite{watt-bcgc} the authors have been
extended the CGC dipole model by involving the dependence of an
amplitude of an impact parameter. Therefore, the dependence of the
saturation scale on impact parameter in expression (8) can be
changed by:
\begin{eqnarray}
 \label{qs-b}
Q_s\mapsto Q_{s}(x,b)=\left(\frac{x_0}{x}\right)^{\frac{\lambda}{2}}\exp\left(- \frac{b^2}{4\gamma_s B_{CGC}}\right) \text{GeV}.
\end{eqnarray}
In Eq. \eqref{qs-b}, instead of $\upsigma_0$ in the CGC dipole model, the free parameter is $B_{CGC}$. It was determined by other reactions, for instance, the $t$-distribution of the exclusive diffractive processes at HERA.

In order to compute the informational entropy associated to the
Color Glass Condensate, let us start by calculating the spatial
Fourier transform of the total deep inelastic cross section, and
thus the color dipole-proton amplitude, is defined in b-CGC model
\cite{watt-bcgc}. For such a purpose, Eqs. (\ref{34} - \ref{333})
are used in the 2-dimensional case. First of all, the Fourier transform
of the total deep inelastic cross section (\ref{CA4}) together with
Eq. (\ref{CA5}) can be computed using Eq. (\ref{34}). After it, the
result of the calculations is used to estimate the modal fraction, Eq.
(\ref{modall}). At last, using an Eq. (\ref{333}) one can calculate
the informational entropy utilizing the modal fraction.

The numerical results follows, after awkward algebraic manipulations. 
The results were obtained for three given values of impact parameter, 
equal to $b = 0$; $b=0.404$, and $b=0.693$, for $x=10^{-4}$ (Fig. 1) and $x=10^{-5}$
(Fig. 2), respectively. As one can see from Fig. 1, the black dotted curve
that depicts the minimum of the nuclear configurational entropy for $b = 0$
corresponds to the value of parameter $c=4.100$. In the case of
impact parameter $b=0.404$ (gray dotted curve), the nuclear configurational entropy
minimum occurs at $c=4.050$ and in the more peripheral collision
when $b =0.693$, which is shown by light gray curve on Fig.1, the
minimum of the informational entropy falls at $c=4.120$,
establishing the onset of quantum regime in the CGC.

In the case of the Bjorken variable given by $x=10^{-5}$, which is shown by Fig. 2, the minimum in
the black dotted curve for impact parameter value $b=0$ corresponds to
$c=4.080$. When value of the impact parameter turn to the $b=0.404$
(grey curve), the minimum of the nuclear configurational entropy corresponds
to $c=4.040$ and in the more peripheral collision when $b=0.693$,
which is shown by light gray curve on Fig.2, the minimum of the
nuclear configurational entropy falls at $c=4.120$. Therefore, one can
conclude the best results of the impact parameter $b=0.404$, that
corresponds to the fitted parameter $c$, and was assumed to be $c=4.01$, in Ref. \cite{Gaya}, to derive the results therein presented. These results match the one in Ref. \cite{Carvalho}. 

The calculation shows that such it is a natural choice provided by the
analysis of the minima of the nuclear configurational entropy, which derives the universal dipole amplitude
with impact parameter of the collision in the Color
Glass Condensate model setup. The uncertainties of the calculation are upper bound by $\sim 1.23\%$.

\begin{figure}[!htb]
       \centering
                \includegraphics[width=.98\linewidth]{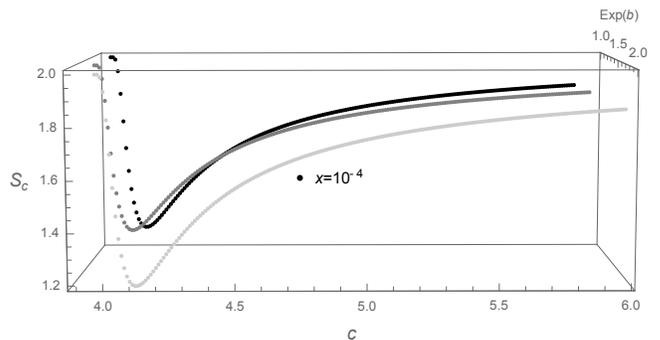}
                \caption{Configurational entropy as a function of the onset of the gluon anomalous dimension, in the Color-Glass Condensate regime,
                 for distinct values of the Bjorken variable $x=10^{-4}$. Each curve was generated for intervals of $c=0.01$. The black curve
is depicted for $b=0$ GeV$^{-1}$, the light grey curve is plot for
$b=0.7$ GeV$^{-1}$ and the grey curve represents $b=0.4$ GeV$^{-1}$.
}
                \label{curv-ricci}
\end{figure}

\begin{figure}[!htb]
       \centering
               \includegraphics[width=.98\linewidth]{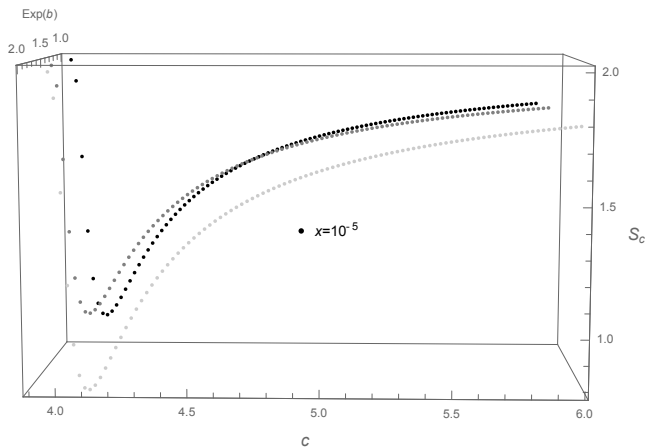}
                \caption{Configurational entropy as a function of the onset of the gluon anomalous dimension,
                in the Color-Glass Condensate regime, for distinct values of the Bjorken variable $x=10^{-5}$. Each curve was generated for intervals of $c=0.01$. The black curve
is depicted for $b=0$ GeV$^{-1}$, the light grey curve is plot for
$b=0.7$ GeV$^{-1}$ and the grey curve represents $b=0.4$ GeV$^{-1}$.}
                \label{curv-ricci}
\end{figure}
\subsection{Outlook}
The nuclear configurational entropy was used 
to derive  the onset of the CGC, with impact parameter 
dependence, matching results in the literature with a {very good accuracy} \cite{Gaya,Carvalho}, of arounf $\sim 1\%$. Figs. 1 and 2, and the analysis that respectively follow, illustrate the 
critical points of the nuclear configurational entropy as 
a way to derive the onset of the quantum regime in the CGC, for different values of the impact parameter $b$. 
A direction to be further studied encompasses quantum mechanics fluctuations. The wave function used
in those equations can be explored in the context of topological defects proposed, e. g., in Refs. \cite{Correa:2016pgr,Correa:2015vka,Bernardini:2016rgb,Bernardini:2016uhj,Bazeia:2013usa,Bazeia:2012qh,Bernardini:2012bh}, in the configurational entropy setup.
{In order to improve the understanding of the nature of  deep inelastic scattering}, as well as the diffractive processes, further systematically investigation is needed concerning
the gluon saturation in the proton wave function.

\acknowledgements
GK thanks to FAPESP (grant No. 2016/18902-9), for partial financial support.


\begin{thebibliography}{99}
\bibitem{Gaya} {}{Karapetyan G.}, \emph{EPL} {\bf 117} {(2017)} {118001}. 
\bibitem{Bernardini:2016hvx} {}{Bernardini A. E. \and da Rocha R.}, \emph{Phys.\ Lett.\ B} {\bf 762} {(2016)} {107}.
\bibitem{Bernardini:2016qit} {}{Bernardini A. E., Braga N. R. F. and da Rocha R.}, \emph{Phys.\ Lett.\ B} {\bf 765} {(2017)} {81}.
\bibitem{Casadio:2016aum} {}{Casadio R. and da Rocha R.}, \emph{Phys.\ Lett.\ B} {\bf 763} {(2016)} {434}.
\bibitem{ads} {}{Braga N. R. F.  and da Rocha R.}, \emph{Phys. Lett. B} {\bf 767} {(2017)} {386}.
\bibitem{Gleiser:2011di} {}{Gleiser M. and Stamatopoulos N.}, \emph{Phys.\ Lett.\ B} {\bf 713} {(2012)} {304}.
\bibitem{Gleiser:2015rwa} {}{Gleiser M. and Jiang N.}, \emph{Phys.\ Rev.\ D} {\bf 92} {(2015)} {044046}.
\bibitem{Gleiser:2012tu} {}{Gleiser M. and Stamatopoulos N.}, \emph{Phys.\ Rev.\ D} {\bf 86} {(2012)} {045004}.
\bibitem{Sowinski:2015cfa} {}{Gleiser M. and Sowinski D.}, \emph{Phys.\ Lett.\ B} {\bf 747} {(2015)} {125}.
\bibitem{Song:2008hj} 
  {}{Song H. \and Heinz U. W.}, 
 \emph{ 
  J.\ Phys.\ G} {\bf 36} {(2009)} {064033}.
\bibitem{kkt} {}{Kharzeev D., Kovchegov Y. V. and Tuchin K.},  \emph{Phys.\ Lett.\ B} {\bf 599} {(2004)} {23}.
\bibitem{Carvalho} {}{Carvalho F., Dur\~aes F. O., Gon\c{c}alves V. P. and Navarra F. S.}, \emph{Mod.\ Phys.\ Lett.\ A} {\bf 23} {(2008)} {2847}. 
\bibitem{cgc} {}{McLerran L. and Venugopalan R.}, \emph{Phys.\ Rev.\ D} {\bf 49} {(1994)} {2233}.
\bibitem{Carvalho:2008zzb} {}{Carvalho F., Dur\~aes F. O., Navarra F. S. and Gon\c{c}alves V. P.}, \emph{Acta Phys.\ Polon.\ B} {\bf 39}{ (2008)} {2511}. 
 \bibitem{Alves:2017ljt}
  {}{Alves A., Dias A. G. \and da Silva R.}, 
  \emph{arXiv:1703.02061 [hep-ph]}.
		\bibitem{Alves:2014ksa}
   {}{Alves A., Dias A. G. \and da Silva R.}, 
  \emph{Physica} {\bf 420} {(2015)} {1}.
\bibitem{watt-bcgc} {}{Watt G. and Kowalski H.},  \emph{Phys.\ Rev.\ D} {\bf 78} {(2008)} {014016}.
\bibitem{pa-R} {}{Rezaeian A. H.}, \emph{Phys.\ Rev.\ D} {\bf 85} {(2012)} {014028}.
\bibitem{ni} {}{Nikolaev N. N. and Zakharov B. G.}, \emph{Z.\ Phys.\ C} {\bf 49} {(1991)} {607}.
\bibitem{gbw} {}{Golec-Biernat K. J. and Wusthoff M.}, \emph{Phys.\ Rev.\ D} {\bf 59} {(1998)} {014017}.
\bibitem{IIM} {}{Iancu E., Itakura K. and Munier S.}, \emph{Phys.\ Lett.\ B} {\bf 590} {(2004)} {199}.
\bibitem{Correa:2016pgr}{}{Correa R. A. C., Dantas D. M., Almeida C. A. S. and da Rocha R.}, 
  \emph{Phys.\ Lett.\ B} {\bf 755} {(2016)} {358}.
\bibitem{Correa:2015vka} {}{Correa R. A. C. and da Rocha R.},  \emph{Eur.\ Phys.\ J.\ C} {\bf 75}{2015}{522}
\emph{Annals\ Phys.}{359} {(2015)} {198}.
\bibitem{Bernardini:2016rgb} {}{Bernardini A. E. and da Rocha R.}, \emph{Phys.\ Lett.\ A} {\bf 380} {(2016)} {2279}.
\bibitem{Bernardini:2016uhj} {}{Bernardini A. E. and da Rocha R.}. \emph{Adv.\ High Energy Phys.} {\bf 2016} {(2016)} {3650632}.
\bibitem{Bazeia:2013usa} {}{Bazeia D., Menezes R. and da Rocha R.}, \emph{Adv.\ High Energy Phys.} {\bf 2014} {(2014)} {276729}.
\bibitem{Bazeia:2012qh} {}{Bazeia D., Losano L., Menezes R. and da Rocha R.}, \emph{Eur.\ Phys.\ J.\ C} {\bf 73} {(2013)} {2499}.
\bibitem{Bernardini:2012bh} {}{Bernardini A. E. and da Rocha R.}, \emph{Adv.\ High Energy Phys.} {\bf 2013} {(2013)} {304980}.







\end{thebibliography}
\end{document}